%% file: main.tex
\documentclass[dvipsnames, twocolumn]{aastex631}

\usepackage{xcolor}
\usepackage{amsmath}
\usepackage{amssymb}
\usepackage{xspace}
\usepackage{enumerate}
\usepackage{acronym}
\usepackage{placeins}
\usepackage{hyperref}
\usepackage[noabbrev,capitalise]{cleveref}
\usepackage{pgf,tikz, bm}
\usetikzlibrary{arrows}

\input{acronyms}
\input{commands}

\begin{document}

\title{Exploring selection biases in \ac{FRB} dispersion--galaxy cross-correlations with magnetohydrodynamical simulations}


\author[0009-0007-8996-0735]{April Qiu Cheng}
\affiliation{Department of Astrophysical Sciences, Princeton University, Princeton, New Jersey 08544, USA}
\affiliation{MIT Kavli Institute for Astrophysics and Space Research, Massachusetts Institute of Technology, 77 Massachusetts Ave, Cambridge, MA 02139, USA}
\affiliation{Department of Physics, Massachusetts Institute of Technology, 77 Massachusetts Ave, Cambridge, MA 02139, USA}

\author[0000-0002-3980-815X]{Shion Elizabeth Andrew}
\affiliation{MIT Kavli Institute for Astrophysics and Space Research, Massachusetts Institute of Technology, 77 Massachusetts Ave, Cambridge, MA 02139, USA}
\affiliation{Department of Physics, Massachusetts Institute of Technology, 77 Massachusetts Ave, Cambridge, MA 02139, USA}

\author[0000-0002-1491-3738]{Haochen Wang}
\affiliation{MIT Kavli Institute for Astrophysics and Space Research, Massachusetts Institute of Technology, 77 Massachusetts Ave, Cambridge, MA 02139, USA}
\affiliation{Department of Physics, Massachusetts Institute of Technology, 77 Massachusetts Ave, Cambridge, MA 02139, USA}

\author[0000-0002-4279-6946]{Kiyoshi W. Masui}
\affiliation{MIT Kavli Institute for Astrophysics and Space Research, Massachusetts Institute of Technology, 77 Massachusetts Ave, Cambridge, MA 02139, USA}
\affiliation{Department of Physics, Massachusetts Institute of Technology, 77 Massachusetts Ave, Cambridge, MA 02139, USA}

\begin{abstract}
The \ac{DM} of \acp{FRB} can be used as powerful probes of the distribution of extragalactic plasma. With a large enough sample, the free-electron--galaxy power spectrum $P_{eg}$ can be measured by cross-correlating \ac{FRB} \acp{DM} with galaxy positions. However, a precise measurement of $P_{eg}$ requires a careful investigation of selection effects: the probability of both observing the \ac{FRB} \ac{DM} and obtaining a host galaxy redshift depends on their properties. We ray trace through the magnetohydrodynamic simulation IllustrisTNG to investigate the impact of expected observational selection effects on \ac{FRB} dispersion--galaxy angular cross-correlations with a sample of 3000 \acp{FRB} at $0.3\leq z\leq 0.4$ . Our results show that cross-correlations with such an \ac{FRB} sample are robust to properties of the \ac{FRB} host galaxy: this includes \ac{DM} contributions from the \ac{FRB} host and optical follow-up selection effects. We also find that such cross-correlations are robust to DM-dependent and scattering selection effects specific to the CHIME/FRB survey. However, a DM-dependent selection effect that cuts off the 10\% most dispersed \ac{FRB} at a fixed redshift shell can bias the amplitude of the cross-correlation signal by over 50\% at angular scales of $\sim 0.1^\circ$, corresponding to $\sim$ Mpc physical scales. Our findings highlight the importance of both measuring and accounting for selection effects present in existing \ac{FRB} surveys, as well as mitigating DM-dependent selection effects in the design of upcoming \ac{FRB} surveys aiming to probe large-scale structure with FRBs.

\end{abstract}



\section{Introduction} \label{sec:intro}
Fast Radio Bursts (FRBs) are extragalactic millisecond-duration radio transients originating at cosmological redshifts. They are characterized by their large dispersive delays, parameterized via \acp{DM}, where the \ac{DM} is directly proportional to the integrated free electron density along the line of sight. Since the majority of the Universe's baryonic matter is in the form of a diffuse plasma outside of galaxies, \ac{FRB} dispersion is an excellent tracer of baryons \changed{\citep{Lorimer_2007, review_2022}}. Unlike other tracers of baryonic matter, such as X-ray observations \changed{\citep[e.g.][]{pen1999, 2018Natur.558..406N}} and the thermal Sunyaev-Zeldovich effect \changed{\citep[e.g.][]{2019MNRAS.483..223T, 2019A&A...624A..48D}}, which have a \changed{signal that is strongly temperature-dependent}, \changed{\ac{FRB} \acp{DM} are sensitive to every free electron along the line of sight}. As such, \acp{FRB} have emerged as promising cosmological probes that can help solve a range of astrophysical problems that are sensitive to the distribution of baryonic matter \citep{Lorimer_2007,Thornton_2013, Mcquinn_2014,Macquart_2020}, from improving galaxy-cluster kinetic SZ measurements \citep{Madhavacheril_2019} to constraining models of astrophysical feedback \citep{Medlock_2024ApJ,Sharma_2025}. 



While to date thousands of \acp{FRB} have been detected \citep{Cat1}, only $\sim 100$  have been localized with the angular resolution necessary to identify their host galaxies and obtain redshifts \changed{\citep{DSA_2019,Heintz_2020,Bhandari_2020, Gordon_2023,Law_2024,Bhardwaj_2024,Sharma_2024,KKO_2025}}. Nevertheless, enough \ac{FRB} redshifts have been measured to enable studies of the extragalactic electron density distribution through the \ac{FRB} line of sight via ``one-point" statistics of \ac{DM} distributions. This typically requires partitioning the total dispersion measure of an \ac{FRB} into its constituent parts (e.g. a contribution from the \ac{FRB} host galaxy, a cosmic contribution from intervening halos and the intergalactic medium, and a contribution from the Milky Way) \citep{Connor_2024, Sharma_2025} that assume a parameterization for each component informed by observations or hydrodynamical simulations. \changed{Parameters of interest can then used to probe a given astrophysical process. 
For instance, the so called “F-parameter”, originally used by \cite{Macquart_2020} to constrain the fraction of all baryons in collapsed halos, parametrizes the variance of the cosmic contribution of the \ac{DM} at a given redshift as $\sigma_{\text{DM, cosmic}}(z)=Fz^{-1/2}$. \cite{Connor_2024} and \cite{Sharma_2025} use simulation-based inference to constrain the strength of different astrophysical feedback processes by modelling the \ac{DM} distribution at a given $z$ as a log-normal.}

In general there are two main disadvantages to using one-point statistics of the DM. The first is that the precision of the DM-budgeting procedure is limited by the accuracy of the (likely overly simplistic) underlying parametrization of each \ac{DM} component, which can be motivated but not verified by simulations. The second is sensitivity to selection effects---the fact that the probability of detecting an \ac{FRB} depends on its observed properties \changed{\citep{2022MNRAS.510L..18J,2022MNRAS.509.4775J}}---which can bias the baryon distribution as measured from \ac{FRB} DMs. For instance, both the ASKAP and CHIME/FRB search pipelines are insensitive to over 50\% of the expected \ac{FRB} population at \ac{DM} $>$ 1000 pc/cm$^{-3}$ \citep{Shannon_2018,Shin_2023, Merryfield_2023AJ}. If, as a simple example, this preferentially removes \ac{FRB} detections that intersect halos with a dense electron density, a measurement of the distribution of baryons in collapsed halos using the one-point statistic would be an underestimate. 

In contrast, ``two-point statistics" of \ac{FRB} \acp{DM} spatially correlated with another tracer of matter such as galaxies \changed{\citep{Masui_2015, Shirasaki_2017, Madhavacheril_2019, Masoud_2020, Masoud_2021, Alonso_2021, Shirasaki_2022,2024PhRvD.110f3556S}}, weak lensing \citep{2023arXiv230909766R}, or even themselves \citep{2023MNRAS.524.2237R,2024PhRvD.110f3556S} are expected to be more robust to selection effects, since the tracers are subject to different observational selection effects that generally do not correlate. Moreover, in the case of cross-correlations with foreground galaxies, only the component of the \ac{DM} that is co-located with the intervening halos \changed{contributes to the expected value of the} measured electron-galaxy power spectrum $P_{eg}(z_g)$ (where $z_g$ is the redshift of the foreground galaxies), allowing it to be naturally extracted without any assumptions on the analytical form of different contributions to the DM. This technique has been demonstrated with $\sim$hundreds of \acp{FRB} by \citet{Connor_ravi_2022, Wu_2023} and later with $\sim$thousands of \acp{FRB} by Wang et al.~(2025, in prep.), albeit without \ac{FRB} redshifts.

However, with upcoming next generation radio telescopes such as CHORD, DSA 2000, and BURSTT, $\sim 10^4-10^6$ \acp{FRB} are expected to be localized to their host galaxies in the next decade \citep{Chord_2019, DSA_2019,BURSTT_2022}, allowing for studies on large-scale structure with \ac{FRB} redshifts to be extended from the existing one-point measurements to two-point cross-correlations. 

In light of thousands to tens of thousands of \ac{FRB} redshifts on the horizon enabling increasingly more precise measurements of the extragalactic electron distribution, 
this work aims to investigate the extent to which selection biases in upcoming \ac{FRB} surveys can be expected to limit the precision of the dispersion measure-galaxy cross-correlation using IllustrisTNG. IllustrisTNG is a suite of large-scale gravomagnetohydrodynamical simulations varying in resolution, physics complexity, and the size of its cubic simulation box, enabling the study of cosmological phenomena across different physical scales \citep{Vogelsberger_2014,illustris_tng_clustering,illustris_tng_groups,illustris_tng_release_2019}. We conduct our simulation with IllustrisTNG's largest simulation box size of $300\,\mathrm{Mpc}$
containing the largest number of galaxies, TNG300-1.

In this work, we demonstrate through our ray tracing simulations which selection biases cross-correlations are the most sensitive to,  along with the rough magnitudes of the effects. In Section \ref{sec:methods} we describe the details of our cross-power spectrum estimation and ray tracing simulations, and in Section \ref{sec:frb_survey} we present and discuss the results of our different simulated selection biases. We conclude in Section~\ref{sec:conclusion}. 

\section{Preliminaries and Methods} \label{sec:methods}
\subsection{The DM-galaxy cross-correlation power spectrum}
\label{sec:preliminaries}
The dispersion measure of an \ac{FRB} observed at a comoving distance $\chi$ and sky location $\vec{\theta}$ is given by
\begin{equation} \label{eq:DM}
    \DM(\chi, \changed{{\vec{\theta}}}) = \int_0^{\chi} \,d\chi^\prime \,n_e(\chi^\prime, \changed{{\vec{\theta}}})(1+z(\chi^\prime)),
\end{equation}
where $n_e$ denotes the comoving electron number density and $z$ is the redshift. Therefore, the average electron number density $\bar{n}_e(z)$ gives us $\overline{\DM}(z)$, the expected \ac{DM} of an \ac{FRB} at redshift $z$; this is the well-known Macquart relation \citep{Macquart_2020}. Given a catalog of \acp{FRB}, we can then define the \ac{DM} \changed{perturbation} as
\begin{equation} \label{eq:ideal_d_def}
    d(\vec{\theta}, z) = \DM(\vec{\theta}, z) - \overline{\DM}(z),
\end{equation}
The \ac{DM} overdensity is thus a measurement of plasma overdensities integrated along the \ac{FRB} line of sight. Indeed, we can similarly define the galaxy overdensity field as
\begin{equation} \label{eq:g_def}
    g(\vec{\theta}, z) = \frac{n_g(\vec{\theta}, z) - \overline{n}_g(\vec{\theta}, z)}{\overline{n}_g(\vec{\theta}, z)},
\end{equation}
where $n_g(\vec{\theta}, z)$ is the observed galaxy number density at $(\vec{\theta}, z)$, and $\overline{n}_g(\vec{\theta}, z)$ is the expected number density in the absence of clustering, which can be a function of sky position if there are non-uniform survey selection effects. 

Given the \ac{DM} and galaxy count overdensity fields, the quantity we are interested in measuring is the angular cross-power spectrum, defined as
\begin{equation} \label{eq:ang_spec_def}
   \delta_{l l^\prime} \delta_{m m^\prime} C^{Dg}_{\ell}(z_{host}, z_g) = \langle \tilde{d}^*_{lm}(z_{host}) \tilde{g}_{l^\prime m^\prime }(z_g) \rangle,
\end{equation}
where $\langle \ldots \rangle$ denotes an ensemble average, and $\tilde{d}_{lm}(z_{host})$ and $\tilde{g}_{lm}(z_g)$ correspond to the \ac{SHT} of the \ac{DM} and galaxy number overdensity fields at their respective redshifts (we will suppress the redshift dependence for simplicity in our following discussion). 
The cross-correlation $C^{Dg}_{\ell}$ between the \ac{DM} field of background \acp{FRB} and a foreground galaxy field at redshift $z_g$ is significant because it provides a direct measuremment of the galaxy-electron 3D power spectrum $P_{eg}$ via the relation \citep{Madhavacheril_2019}
 \begin{equation}
 \label{eqn:CDG_PEG}
     C^{Dg}_{\ell} = n_{e,0}\frac{1+z_g}{\chi_g^2}P_{eg}\left(k=\changed{\frac{\ell+1/2}{\chi_g}},z_g\right),
 \end{equation}
$P_{eg}$, in turn, probes the distribution of baryonic matter around galaxies. \changed{Note that \cref{eqn:CDG_PEG} is derived using the Limber approximation, which is valid up to very large scales (small $\ell$), at which we are simultaneously limited by the finite simulation periodic box size.}

\subsection{Power Spectrum Estimation}
In practice, we discretize the survey volume, and quantities $\DM$, $n_g$, etc.\ are evaluated by binning the \acp{FRB} and galaxies into angular sky pixels and tomographic redshift bins. Note that an \ac{FRB} survey only yields sparse measurements of the \ac{DM} field at the sky locations of the survey \acp{FRB}. That is, $\DM(\vec{\theta}, z)$ only has a measured value in pixels that contains at least one \ac{FRB}; an average is taken if there is more than one \ac{FRB} in a pixel.

Our mock catalogs examine the sky in square patches of approximately $6^\circ \times 6^\circ$, a choice required by the geometry of the simulation box (see Appendix~\ref{app:skypatch} for a detailed discussion). Therefore, we can use the flat sky approximation, in which the \ac{SHT} reduces to a 2D \ac{DFT}: 
\begin{equation} \label{eq:DFT}
    \tilde{f}(\vec{\ell}) = \frac{A}{N^2} \sum_{x=0}^{N-1} \sum_{y=0}^{N-1} f(\vec{\theta}_{x, y}) \, \exp(-i \, \vec{\ell}  \cdot \vec{\theta}) 
\end{equation}
for some field $f(\vec{\theta})$, where $A$ is the angular area of the sky patch and $N$ is the number of pixels on a side.
The flat sky angular power spectrum is then
\begin{equation} \label{eq:ang_spec_def_flat}
    \delta_{\vec{\ell} \vec{\ell}^\prime} C_l^{Dg} = \frac{1}{A}\langle \tilde{d}^*_{\vec{\ell}} \,\tilde{g}_{\vec{\ell}^\prime} \rangle.
\end{equation}
Naively, the power spectrum above can be estimated by averaging over the independent Fourier modes\footnote{For any real field $f$, the reality condition imposes that $\tilde{f}_{\vec{\ell}} = \tilde{f}^*_{-\vec{\ell}}$.} $\vec{\ell}$ that satisfy $|\vec{\ell}| = \ell$:
\begin{equation} \label{eq:naive_estimator_flat}
     \hat{C}_{\ell}^{Dg}  =\frac{1}{A N_\ell} \sum_{|\vec{\ell}| = \ell} \tilde{d}^*_{\vec{\ell}} \,\tilde{g}_{\vec{\ell}},
\end{equation}
or, if we bin into bandpowers $\lambda$,
\begin{equation} \label{eq:naive_estimator_flat_binned}
     \hat{C}_{\lambda}^{Dg}  =\frac{1}{A N_\lambda} \sum_{|\vec{\ell}| \in \lambda} \tilde{d}_{\vec{\ell}} \,\tilde{g}^{*}_{\vec{\ell}},
\end{equation}
where $N_\lambda$ is the total number of independent modes belonging to bandpower $\lambda$. 

However, in any realistic survey, we will not have an \ac{FRB} in every sky pixel, and thus the \ac{DM} measurement is incomplete. This is equivalent to multiplying the \ac{DM} field with a boolean window function $w^d_{\vec{\theta}}$ corresponding to whether or not at least one \ac{FRB} was detected at $\vec{\theta}$. Similarly, the galaxy field can have its own survey window $m^g_{\vec{\theta}}$. With incomplete surveys, the naive estimator defined in \cref{eq:naive_estimator_flat} can only operate on the masked fields, namely
\begin{equation} \label{eq:naive_estimator_flat_mask}
     \hat{C}_{\lambda}^{Dg} =\frac{1}{A N_\lambda} \sum_{|\vec{\ell}| \in \lambda} (\tilde{d}^m_{\vec{\ell}})^* \,\tilde{g}^{m}_{\vec{\ell}},
\end{equation}
where $d^m_{\vec{\theta}} = m^d_{\vec{\theta}}\,d_{\vec{\theta}}$ and $g^m_{\vec{\theta}} = w^g_{\vec{\theta}}\,g_{\vec{\theta}}$ are the masked \ac{DM} and galaxy overdensity fields.
In this case, the estimated power spectrum usually suffers from mixing of power across different angular scales, i.e.,
\begin{equation} \label{eq:naive_estimator_problem}
    \hat{C}^{Dg}_{\lambda} = \sum_{\lambda^\prime} M_{\lambda \lambda^\prime} \overline{C}^{Dg}_{\lambda^\prime},
\end{equation}
where $\overline{C}^{Dg}_{\lambda^\prime}$ is the true underlying angular power spectrum, and $M_{\lambda \lambda^\prime}$ is the power mixing matrix which depends on the survey windows $m^d$ and $m^g$.

To obtain an unbiased estimate for the cross-power spectrum, we can use the \ac{OQE} derived in \cite{tegmark_1997}:
\begin{equation} \label{eq:oqe}
    C_\lambda^{Dg} = \sum_{\lambda^\prime} (\fisher)^{-1} \,\frac{(\mathbf{g}^m)^\dagger \icovg \dcovp \icovd \mathbf{d}^m}{2}.
\end{equation}
Here, $\mathbf{g}^m$ and $\mathbf{d}^m$ are the masked real space galaxy and \ac{DM} overdensity fields flattened into vectors and $\mathbf{D} = \langle \mathbf{d}^m (\mathbf{d}^m)^\dagger\rangle$, $\mathbf{G} = \langle \mathbf{g}^m (\mathbf{g}^m)^\dagger \rangle$, and $\cov = \langle \mathbf{d} \mathbf{g}^\dagger \rangle$ are the \ac{DM}, galaxy, and \ac{DM}-galaxy (cross-)covariance matrices, respectively. $\dcovp = \partial\cov/\partial\ensuremath{C_{\lambda^\prime}^{Dg}}$ is the derivative cross-covariance matrix, which can be written in terms of the DFT operators on the masked fields and whose form we derive in Appendix~\ref{app:estimator}. \fisher~is the Fisher information matrix, which here is
\begin{equation} \label{eq:fisher}
    \fisher = \frac{1}{2} \trace [ \dcov \icovg \dcovp  \icovd ].
\end{equation}
In a realistic survey, the full covariances $\mathbf{D}$ and $\mathbf{G}$ can be difficult to estimate. Even if they can be obtained, the computational cost for \cref{eq:fisher} is often too high. Works such as \citet{Alonso_2019} have shown that simplifications by assuming the covariances $\icovd$ and $\icovg$ are diagonal allow \cref{eq:oqe,eq:fisher}  to be computed efficiently while keeping the estimator close to optimal in most cases. One simple way to estimate $\icovd$ and $\icovg$ is to have some estimated variance $\sigma^{-2}$ along the diagonal for pixels with at least one \ac{FRB}, and $0$ for pixels with no \acp{FRB}, corresponding to a limit of infinite noise (i.e. the \ac{DM} is not measured). However, any constant that multiplies $\icovd$ or $\icovg$ will cancel out between the inverse Fisher matrix and the numerator in \cref{eq:oqe}. Therefore, we can simply take the diagonal of \icovd and \icovg to be the survey mask:
\begin{align}
    \icovd = \text{diag}(\mathbf{m}^d) \label{eq:D_inv} \\
    \icovg = \text{diag}(\mathbf{m}^g), \label{eq:G_inv}
\end{align}
where $\mathbf{m}^d$ and $\mathbf{m}^g$ are the \ac{FRB} survey mask and galaxy survey mask flattened into 1-D vectors, respectively. 
As a sanity check, we show in Appendix~\ref{app:estimator} that this estimator reduces to the naive estimator in \cref{eq:naive_estimator_flat_binned} in the limit of complete sky coverage.

Finally, the error on the \ac{OQE}, assuming the fields are Gaussian and that there is no window function present, can be shown to be \citep{tegmark_1997}
\begin{equation} \label{eq:theory_variance}
    \langle (C_\lambda^{Dg} - \overline{C}_\lambda^{Dg})^2\rangle = (\Delta C_\lambda)^2 = \frac{\overline{C}^{DD}_\lambda \overline{C}^{gg}_\lambda + \left( \overline{C}^{Dg}_\lambda \right)^2}{N_\lambda},
\end{equation}
where $C^{DD}_\lambda$ and $C^{gg}_\lambda$ are the auto power spectrum for the \ac{DM} and galaxy overdensity field, respectively. However, as we discuss further in Section \ref{subsec:fiducialpower}, the Gaussianity assumption on which this equation relies is unlikely to be valid on small scales due to non-linear structure growth.  We therefore estimate the errors of our cross-power spectrum empirically, by computing it over multiple realizations in many independent sky patches (see Section~\ref{sec:sim}).



\subsection{Ray Tracing with IllustrisTNG}\label{sec:sim}
The IllustrisTNG project is a suite of large-scale, cosmological magnetohydrodynamical simulations that model galaxy formation and evolution across cosmic time.
IllustrisTNG consists of three volumes with box sizes 50, 100, and 300 cMpc---where c denotes comoving distance--enabling cosmological studies focusing on different scales; each box size is run at three different resolution levels.
For our study, we choose TNG300-1, the highest resolution run of the largest simulation box in order to simulate \acp{FRB} found at cosmological distances. 
Each simulation comes in 100 discrete ``snapshots'' containing field data\footnote{See \url{https://www.tng-project.org/data/docs/specifications}.} of all particles in the simulation box, of which 80 are ``mini" snapshots that contain a subset of the field particle information as the ``full" snapshots. 
We access the data via the online IllustrisTNG JupyterLab workspace\footnote{\url{https://www.tng-project.org/data/lab}}.

The size of such a cosmological simulation is extremely large. The simulation box of TNG300-1 consists of $2500^3$ particles, with a total size of 4.1 TB per full snapshot. Directly ray tracing through the simulation data would require a linear search of all particles in the simulation box along the given sightline, a computationally infeasible task for even a single \ac{FRB}. Previous works \citep{Zhang_2021,Walker_2024} have compressed the data by choosing a single direction along which to ray trace and pre-computing the \ac{DM} along those sightlines. 
Because we need to be able to place \acp{FRB} in arbitrary directions, we instead compress the data by computing the electron density field of each snapshot on a grid. 
To do so, we sort the particles of a given snapshot into cubic bins.
The electron number count of each particle is then computed as
\begin{equation}
    N_e = \eta_e\,X_H\frac{\rho}{m_p}V
\end{equation}
where $\eta_e$ is the electron abundance with respect to the hydrogen number density, $X_H$ is the total hydrogen abundance, $\rho$ is the comoving mass density, $m_p$ is the proton mass, and $V$ is the volume of the particle. 
The comoving electron density of each bin is obtained by dividing its total electron number count by its volume. We then repeat this for each snapshot. Note that while $X_e$ and $\rho$ are available in all snapshots, $X_H$ is only available in the full snapshots, and we take $X_H = 0.76$ for the mini-snapshots. We have checked and found no discontinuities in the line-of-sight electron density as a result of this approximation.

Due to the memory limitations of the JupyterLab workspace, which limits the size (total number of pixels) of the output gridded map that can be held in memory, we choose a bin resolution of $500\,\ckpch$. We note that our pixel resolution of 500 ckpc/$h$ places an upper limit of $k \lesssim \sim 6\,h\,\mathrm{Mpc}^{-1}$ on the scales at which we can resolve $P_{eg}$, \changed{whereas the finite periodic box size sets a lower limit of $k \gtrsim 0.02$. For our fiducial $0.2 < z_g < 0.3$, this corresponds to $20 \lesssim \ell \lesssim 5000$.}

Another difficulty in ray tracing in IllustrisTNG comes from the fact that \acp{FRB} are generally observed to come from cosmological distances ($z\gtrsim0.1$) \citep{Macquart_2020} that exceed the $300\,\mathrm{cMpc}$ size of the simulation box. Since the simulation box has periodic boundary conditions, given observer and \ac{FRB} positions, computing the \ac{FRB} \ac{DM} can be done by stacking simulation snapshots. However, this raises the possibility of a sightline (or nearby sightlines within the same sky region) intersecting the same part of the simulation box multiple times. Previous works have dealt with this issue by either changing the angle of the line of sight segment or performing linear coordinate transformations (e.g. rotating the box) when the ray crosses the box boundary \citep{Zhang_2021,Walker_2024}. While this solution is sufficient for studies of \ac{DM} distributions, the cross-correlation is a statistic that explicitly describes spatial correlations, and therefore requires the preservation of relative positions of particles within the light cone.  We resolve this problem geometrically, computing a-priori the ``good" regions of sky, where no sightline intersects with itself or its neighboring pixels for a given pixel resolution. 

\begin{figure}[t]
\includegraphics[width=\columnwidth]{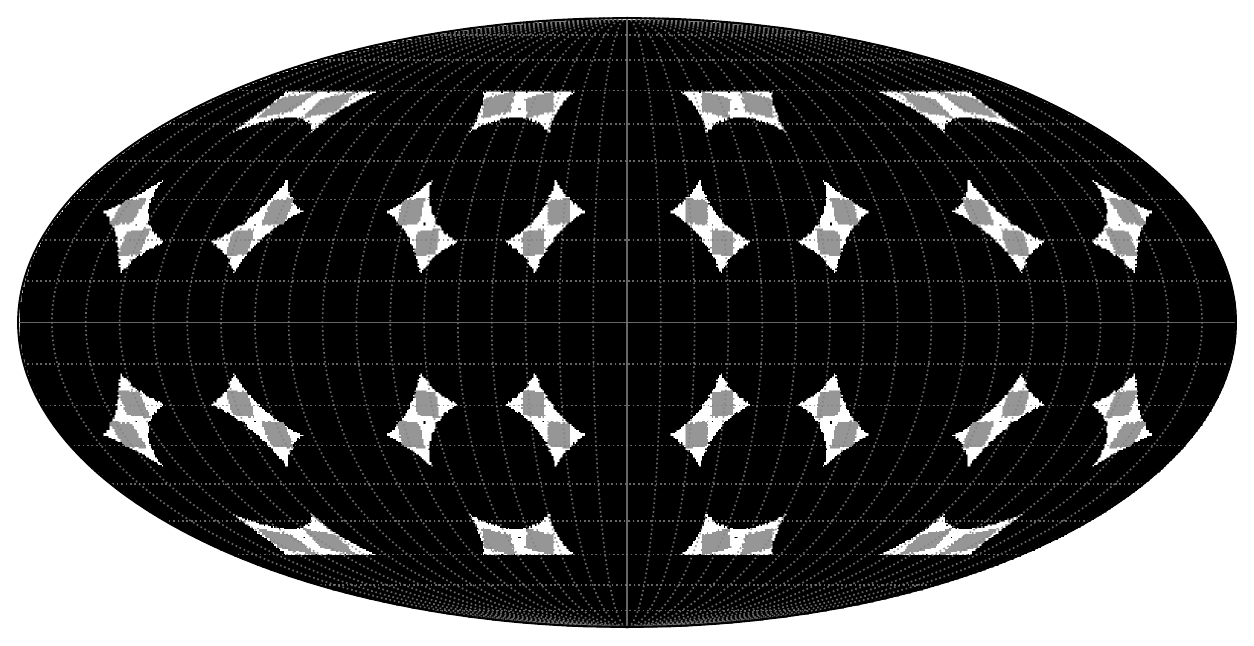}
\caption{The $48$ square sky regions (shaded gray) used in this study (out to $z_{\max} = 0.4$) for independent realizations of the \ac{DM} and galaxy field. The regions are chosen from the ``good" regions of the sky (white) that avoid spurious geometrical effects from the simulation's periodic boundary conditions.}
\label{fig:regions}
\end{figure}

The larger the distance to a given \ac{FRB}, the more times the simulation box must be stacked and therefore the more likely it is that a given line of sight will intersect with the same structure more than once. Equivalently, the larger the \ac{FRB} redshift, the smaller the allowed sky regions become, therefore requiring a compromise between these quantities. We leave the details of these computations to Appendix \ref{app:skypatch}. The result is summarized in \cref{fig:regions}, which shows the allowed sky regions for our choice of $z_{\max}=0.4$ and pixel resolution $0.0008\,\mathrm{rad}$, along with the $48$ $0.1\,\rad \times 0.1\,\rad$ square sky patches within the allowed regions. There are $125^2 = 15625$ pixels per sky region.


Finally, our ray tracing procedure can be described as follows. Given any observer and \ac{FRB} position, we:
\begin{enumerate}
    \item Compute the total path of the ray through the box by evolving the ray from the observer to the \ac{FRB}. 
    \item For each cubic bin the ray intersects, retrieve its electron density from the closest snapshot. The closest snapshot is defined as the snapshot with the smallest comoving distance between the center of the ray within the bin and the redshift at which the snapshot was taken.
    \item Compute and save the cumulative $\DM(\chi)$ with \cref{eq:DM} using trapezoidal Riemann integration:
    \begin{equation}
        \DM(\chi) = \sum_{\chi_i \leq \chi} n_e(\chi_i) \bigg(1+z(\chi_i)\bigg) \Delta \chi_i.
    \end{equation}
    where $\Delta \chi_i$ is the comoving length of the ray in the $i$th cubic bin, accounting for ``partial intersections''.
\end{enumerate}
By computing the cumulative line-of-sight \ac{DM}, we only have to ray trace once for each pixel of each sky region. Then, we can compute the \ac{DM} of an \ac{FRB} placed within a sky region at an arbitrary $z < 0.4$ via interpolation with the radial coordinate.
We choose the same observer position for all sky regions, which is the lowest density region of the simulation box as our origin, since the binning of the simulation box wipes out nearby structures. Effectively, this is an explicit removal of any galactic contribution to the \ac{DM}, corresponding to the assumption that the Milky Way \ac{DM} is known and can be subtracted in a real survey.



Finally, the \ac{DM}-galaxy cross-correlation requires a catalog of foreground galaxies. Because there are far fewer galaxies than simulation particles, it is not necessary to bin the galaxy map as we did for electrons. IllustrisTNG provides galaxies in the form of a ``group catalog" at each snapshots, containing a list of halos and subhalos. Galaxies are taken to be subhalos flagged as galaxies (``\texttt{SubhaloFlag=0}") with nontrivial stellar content ($M_g < 0$). \changed{Note that while this results in a halo mass cutoff that is lower than realistic, imposing realistic apparent magnitude cuts to our mock galaxy catalog was not found to significantly change our results, with only a modest increase in the error bars.} 

\changed{With the same simulation volume as the \ac{DM} map,} we build the light cone of galaxies in a similar fashion as our ray tracing procedure, stacking simulation boxes and evolving the cone in tomographic slices corresponding to each snapshot. 
We identify the redshift range each snapshot, computed as the halfway point (in comoving distance) to neighboring snapshots. 
Then, for each snapshot $i$, we create a galaxy catalog corresponding to that slice by taking all galaxies with $\chi(z_{\min, i}) \leq|\vec{x}_g - \vec{x}_\text{obs}| < \chi(z_{\max, i})$, where $\vec{x}_g$ and $\vec{x}_\text{obs}$ are the positions of the galaxy and observer, respectively. 
We save the galaxies' center-of-mass coordinates, along with other properties retrieved from IllustrisTNG (stellar mass, \ac{SFR}, and $g$-band absolute magnitude), from which we straightforwardly compute the projected sky position and apparent magnitude. The end product is a catalog of galaxies within the cone of each sky region up to a maximum redshift.

Combined, these tools enable generic studies of realistic \ac{FRB} surveys. In particular, we now use this framework to study potential systematics in the \ac{DM}-galaxy cross-correlation in realistic \ac{FRB} and galaxy surveys.

\subsection{Fiducial cross-power spectrum} \label{subsec:fiducialpower}
Using the methods described in the previous two sections, we populate the simulation volume with \acp{FRB}, use our ray-tracing methods to determine their \acp{DM}, and measure \ClDg with our cross-power spectrum estimator. We now describe the parameters of our fiducial population model, without selection effects, and compare the resulting power-spectrum to theoretical expectations.

In all our experiments, we simulate $N=3000$ \acp{FRB} within a redshift slice $0.3 < z_\text{host} \leq \changed{{z_\mathrm{\max}} = }0.4$ for each sky region shown in \cref{fig:regions}. We draw our \acp{FRB} randomly from galaxies weighted by the galaxy's \ac{SFR} under the simple model that \acp{FRB} trace the star formation rate:
\begin{equation} \label{eq:sfrweight}
    W_\text{host} = \frac{\mathrm{SFR}_\text{host}}{1+z_\text{host}}
\end{equation}
where the $1+z_\text{host}$ arises from the cosmological time dilation of the rate. We then cross-correlate the \acp{DM} of our \acp{FRB} with foreground galaxies in a $0.2 < z_g \leq 0.3$ redshift slice.

\begin{figure}
  \includegraphics[width=\columnwidth]{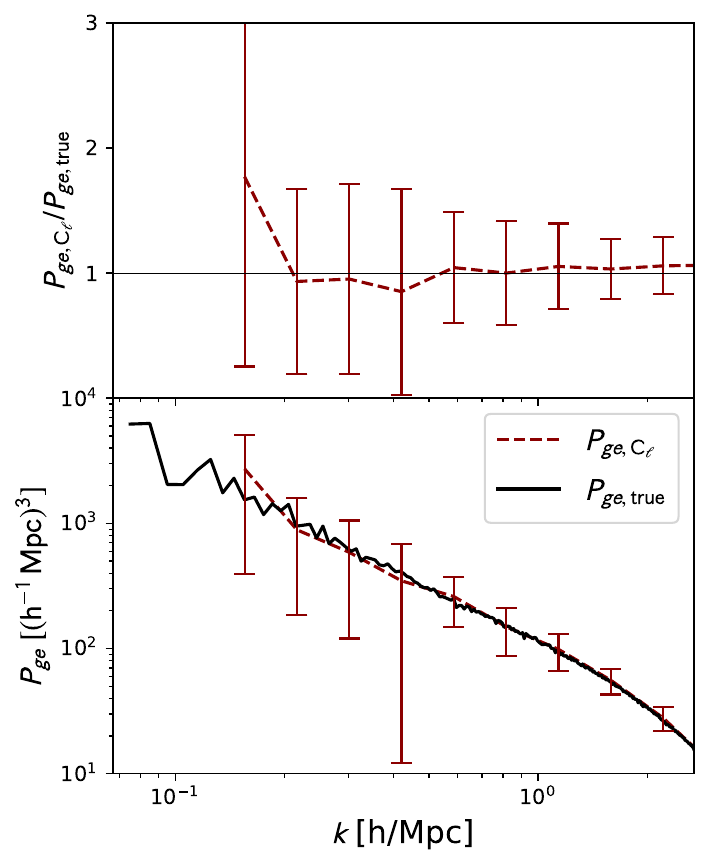}
  \caption{Comparison of $P_{eg}(z_g)$ computed directly from the electron density and galaxy overdensities at redshift $z_g=0.2$ (black) to the electron galaxy power spectrum calculated from $C_\ell^{Dg}$ using Equation \ref{eqn:CDG_PEG} (red). The top panel shows the ratio of the two power spectra, with the error bars calculated from the variance of different realizations of $C_\ell^{Dg}$. This demonstrates the validity of the theory that relates the observable dispersion angular power spectrum to the underlying three dimensional power spectrum of free electrons.
  }
  \label{fig:peg}
\end{figure}

The DM-galaxy cross-correlation ideally correlates the \ac{DM} of a layer of background \acp{FRB} with a layer of foreground galaxies. In this case, \ClDg becomes a direct proxy for the 3D galaxy--electron power spectrum evaluated at the redshift of the foreground layer $P_{eg}(z_g)$, as given by Equation~\ref{eqn:CDG_PEG}. We demonstrate this in Figure \ref{fig:peg},
which shows that the electron-galaxy power spectrum as inferred from our fiducial measurement of \ClDg is in good agreement with the ``true" electron-galaxy power spectrum, obtained by directly measuring $P_{eg}(z_g)$ from the electron density map and halo catalog at redshift $z_g$. As stated earlier, our pixel resolution of 500 ckpc/$h$ introduces a window function that limits the scales at which we can obtain $P_{eg}$, and that beyond $k\sim$ 1 $h\,\mathrm{Mpc}^{-1}$ the pixelation is expected to affect the amplitude of the power spectrum by more than 2$\%$. As such, we only plot our power spectra out to a value of $k\sim 2.6$ $h\,\mathrm{Mpc}^{-1}$ ($\ell\sim4000$). 

We estimate the uncertainties on the cross-correlation empirically; each of the $48$ sky regions represents an independent realization of the matter field, and we have $n_\text{trial}=5$ realizations of the \ac{FRB} survey (via independent draws from galaxies) for each region, for a total of $240$ realizations of \ClDg. All figures show the mean and standard deviation \ClDg from these $240$ total realizations unless otherwise specified.

We also note that the error bars (standard deviation over all $C^{Dg}_\ell$) placed on our power spectra are the error bars corresponding to a survey of 3000 FRBs, not the uncertainties on the reported mean simulated power spectrum measurement (the standard error) which are a factor of $\sqrt{48}$ smaller. The factor of $\sqrt{48}$ can be understood as follows: the total contribution to the cross correlation noise in an ideal survey of \acp{FRB} with redshifts (i.e. the Maquart relation is removed) can be decomposed into (a) sample/cosmic variance (b) \ac{FRB} host galaxy/local environment noise and (c) foreground galaxy shot noise. \changed{As discussed in Section~\ref{sec:sim}, we use a large sample of all foreground galaxies, and thus the galaxy shot noise is negligible}, leaving contributions (a) and (b).  While each of the 5 trials randomizing the choice of background \acp{FRB} reduce the contribution of (b), the large-scale structure sample variance is the same for each trial within a given sky region, meaning we have only 48 effective realizations of (a). 





%

\begin{figure}
`\includegraphics[width=\columnwidth]{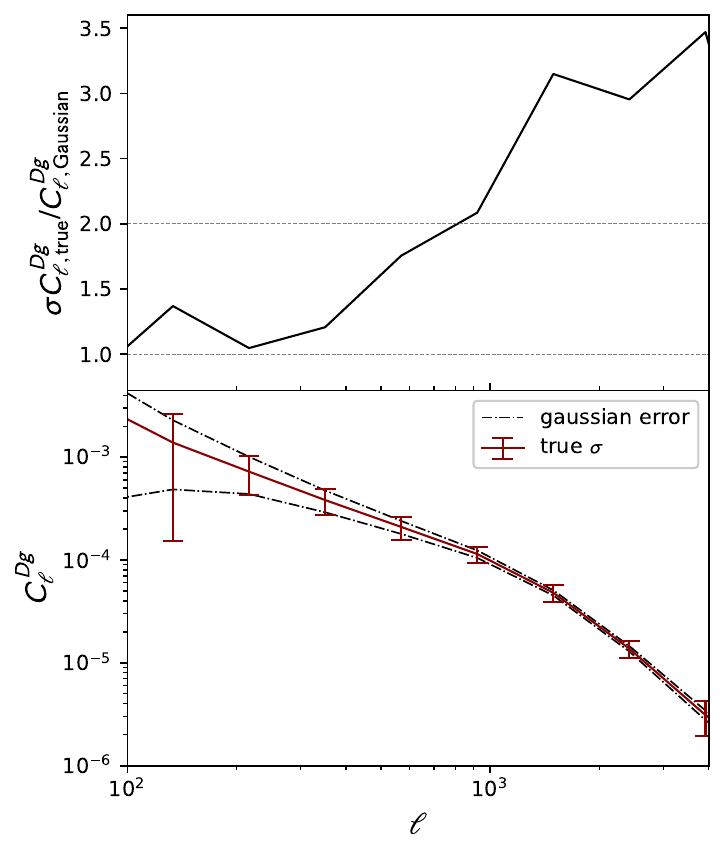}
\caption{Comparison of theoretical expectation of the error on $C_\ell^{DG}$ as predicted by Equation \ref{eq:theory_variance} in the case of a Gaussian field to the true standard deviation of our measured $C_\ell^{DG}$ over many realizations for \acp{FRB} cross correlated with galaxies. Here, we have \acp{FRB} drawn uniformly distributed over the sky (one FRB/pixel) to avoid the additional effect the window function has on the variance. The true variance of the data is increasingly larger than the Gaussian expectation at smaller scales.}
\label{fig:non_gaussianities_vs_gaussian}
\end{figure}

Finally, we note that forecasts (e.g. \cite{Madhavacheril_2019}) often make two central assumptions to obtain theoretical error bars (as given by Equation \ref{eq:theory_variance}) on $C_\ell^{Dg}$: the first is that no window function is present, and the second is that the \ac{DM} and galaxy fields are gaussian. In order to isolate the effect of the second--the non-gaussianity of the cosmic strucutre--we compare the actual standard deviation of the power spectrum for a sample of 15,625 \acp{FRB} (one \ac{FRB} per pixel) to the expectation for Gaussian fields in Figure \ref{fig:non_gaussianities_vs_gaussian}. 
We find that the Gaussian expectation underestimates the true error particularly at large $\ell$, which is a natural consequence of the fact that matter over densities become increasingly non-Gaussian at small scales due to non-linear structure growth. As such, future forecasts of this measurement should adjust error bars accordingly---although in practice, with real data the errors are computed using mock or randomized galaxy and \ac{FRB} catalogs (see e.g. \cite{Masoud_2021}, Wang et al. (2025, in prep.).

\begin{figure}
\includegraphics[width=\columnwidth]{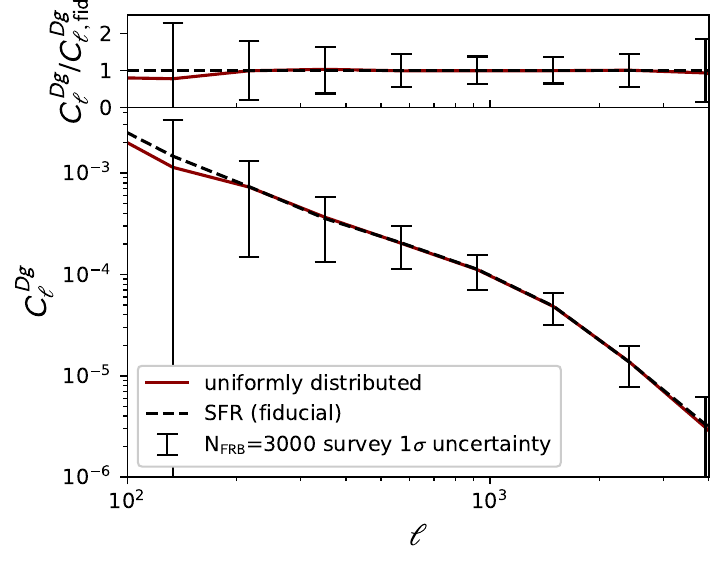}
\caption{Comparison of cross correlation measurement when \acp{FRB} are drawn uniformly over all pixels (red) to when \acp{FRB} are drawn based on SFR (Equation \ref{eq:sfrweight}). That \acp{FRB} stocastically---rather than uniformly---sample the \ac{DM} field induces no bias on the power spectrum.}
\label{fig:sfr_uniform}
\end{figure}

\section{Impact of Selection Effects}  \label{sec:frb_survey}
\subsection{Insensitivity to FRB and host galaxy properties}
\label{sec:host_selection_effects} 






\begin{figure}[h]
\includegraphics[width=\columnwidth]{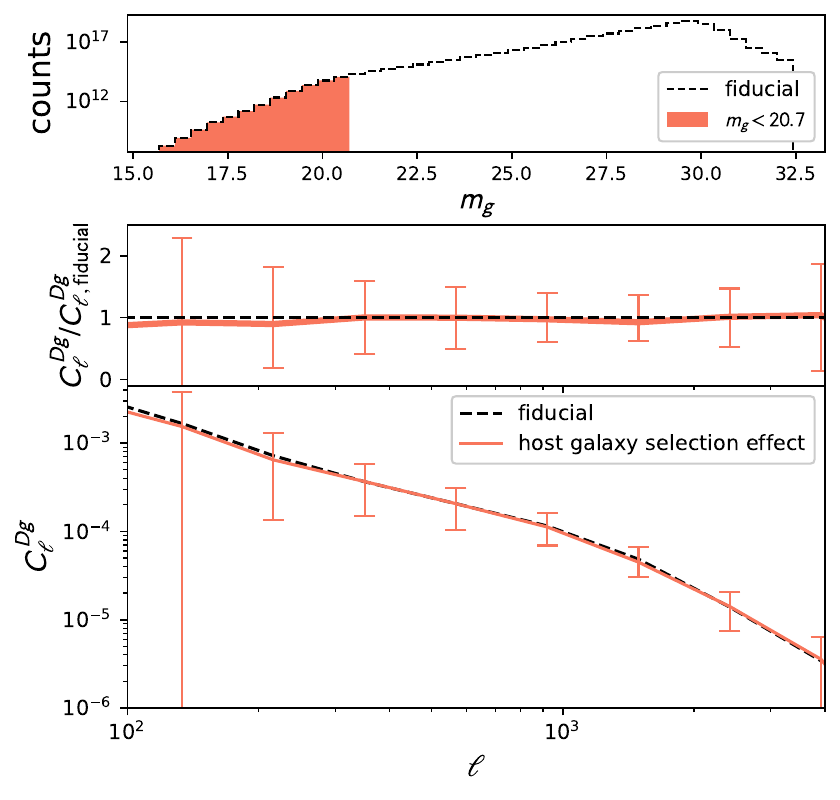} 
\caption{Top: Apparent magnitude distribution for fiducial sample of \acp{FRB} in the redshift range ($0.3 < z_g \leq 0.4$). Bottom: Host galaxy incompleteness does not significantly affect the cross-correlation, even for an aggressive $m_g$ cut. }
\label{fig:mg}
\end{figure}

\begin{figure}
  \includegraphics[width=\columnwidth]{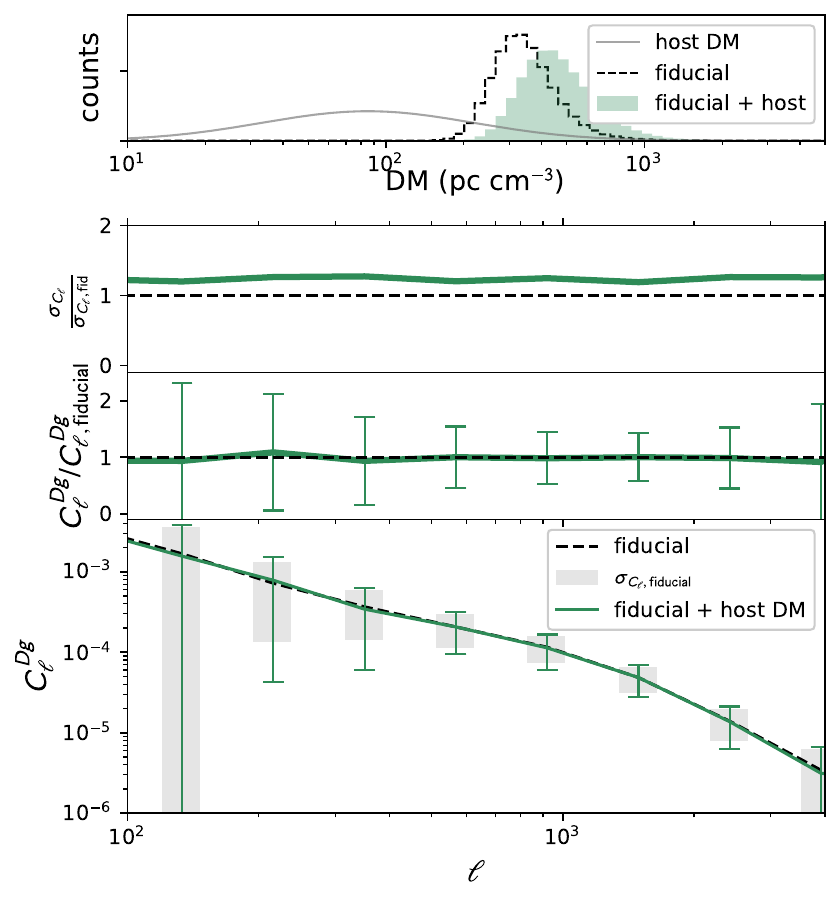}
  \caption{\textit{Top}: The \ac{DM} distribution of our fiducial sample of \acp{FRB} and our sample after injecting a host \ac{DM} contribution. \textit{Bottom}: The cross-correlation with the fiducial model and after injecting a host \ac{DM}. The ratio of the error after injecting a host \ac{DM} to the error on the fiducial model is shown in the second plot from the top, and the ratio of the $C^{Dg}_\ell$s are given in the third plot from the top. While the host \ac{DM} contribution contributes noise to $C_\ell^{Dg}$, it does not significantly bias the cross correlation measurement.}
  \label{fig:hostDM}
\end{figure}

\begin{figure}[h]
\includegraphics[width=\columnwidth]{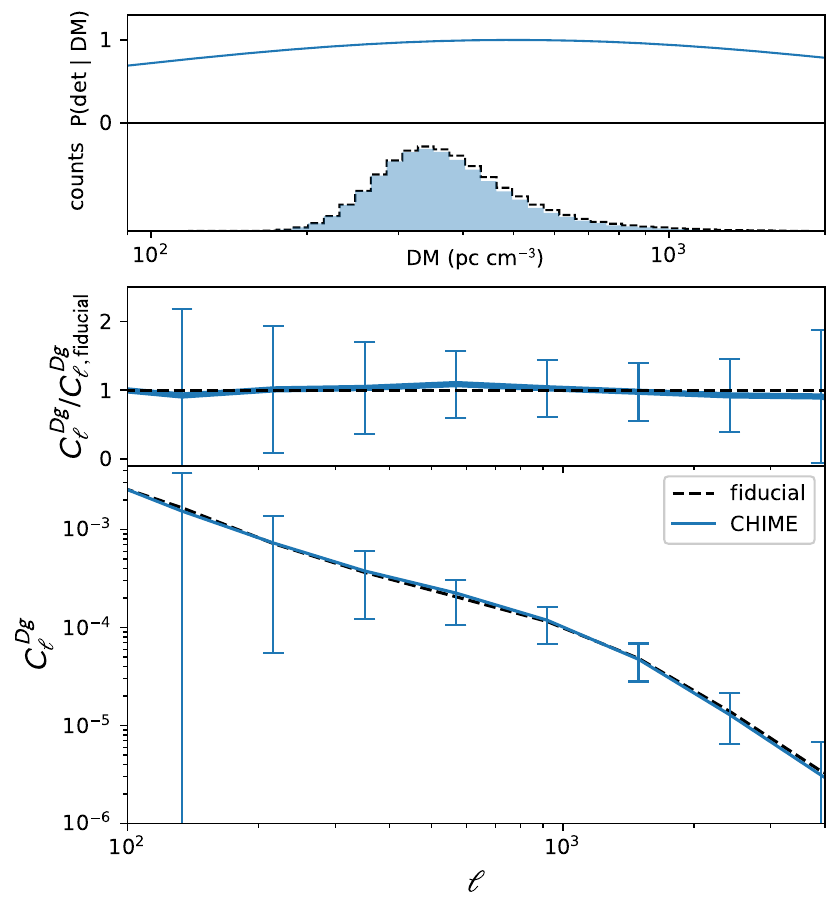}
\caption{\textit{Top}: The change in \ac{DM} distribution after introducing a \ac{DM}-dependent selection function as given by \cref{eq:DM_sfunc} for $a=2$, which roughly scales the DM-dependent selection effect from the CHIME/FRB survey down to our \ac{FRB} redshift range. \textit{Bottom}: The cross-correlation of the \ac{DM} field with the DM-dependent selection effect applied, compared to the default model as reference. The residuals are plotted in the middle panel. A smooth DM-dependent selection effect not significantly affect the cross-correlation signal.}
\label{fig:DM_sfunc1}
\end{figure}

\begin{figure}[h]
\includegraphics[width=\columnwidth]{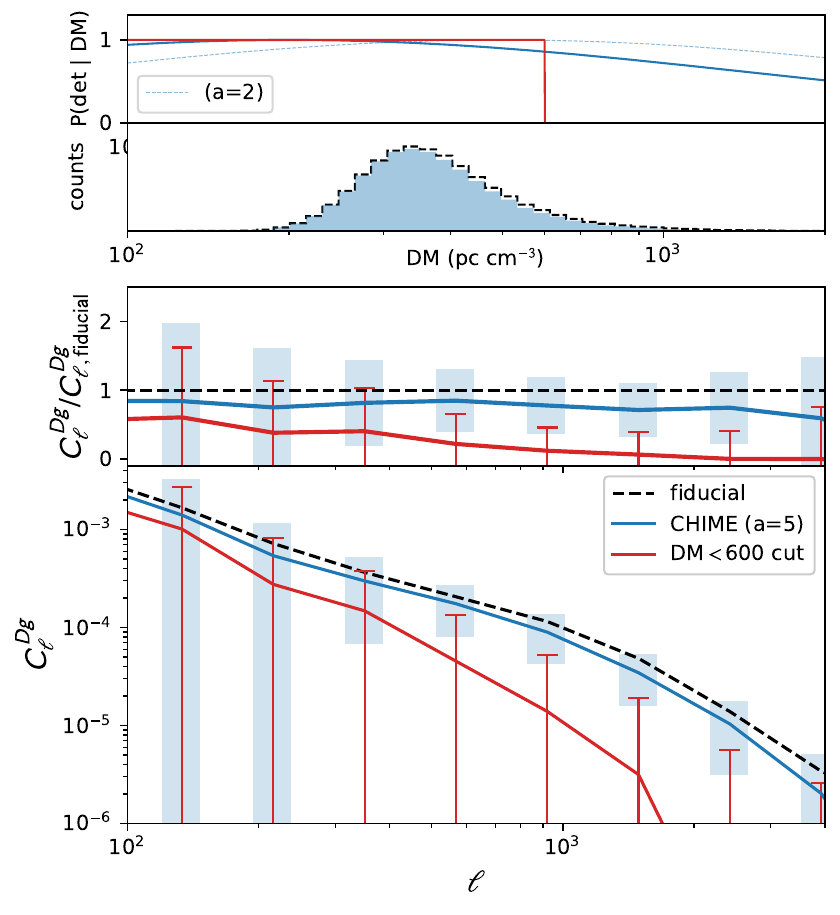}
\caption{\textit{Top}: The change in \ac{DM} distribution after introducing a \ac{DM}-dependent selection function as given by \cref{eq:DM_sfunc} for $a=5$ (solid blue) and a step-wise selection function that cuts off at 800 pc cm$^{-3}$ (red). The DM-dependent selection function for $a=2$ shown in Figure \ref{fig:DM_sfunc1} to represent CHIME's selection effects is plotted with the dashed blue line for reference. \textit{Bottom}: The cross-correlation of the \ac{DM} field with the DM-dependent selection effect applied, compared to the default model. The residuals are plotted in the middle panel. The shaded boxes and error bars indicate the standard deviation for the CHIME-like and step-wise \ac{DM} dependent selection functions, respectively. With a \ac{DM} cut given by \cref{eq:DM_sfunc}  with $a=5$, the amplitude of the power spectra drops by over 20\% beyond $\ell \sim 10^3$. For the abrupt \ac{DM} cut, the amplitude of the power spectra drops by over 50\% beyond $\ell \sim 10^3$.}
\label{fig:DM_sfunc}
\end{figure}

Because, in principle, the signal in the cross-correlation arises purely from the electrons along the \ac{FRB} line of sight that are clustered around foreground galaxies, it is convenient to consider two different kinds of selection effects. The first are selection effects on observational properties of the \ac{FRB} and host galaxy that arise outside of the foreground galaxy redshift shell. The second are selection effects on observational properties of the \ac{FRB} that arise within the galaxy redshift shell, which we will refer to as ``propagation selection effects". In this section, we consider the former and in section \ref{sec:propagation} we consider the latter. 


\subsubsection{FRB population model}
\label{subsec:FRB_Population_Model}
We adopt a distribution of \acp{FRB} that trace the SFR of galaxies within a specified redshift shell in our fiducial model. As discussed in Section \ref{sec:preliminaries}, this introduces a window function on the \ac{DM} overdensity map that mixes power on different angular scales, which can be mitigated with the OQE. We test the effect of the physical placement of \acp{FRB} on the sky in Figure \ref{fig:sfr_uniform} by comparing our  measurement of $C_\ell^{Dg}$ with the fiducial \ac{FRB} sample (criteria listed in Section \ref{subsec:fiducialpower}) to our measurement with an identical sample of \acp{FRB} except with \acp{FRB} uniformly drawn over the sky (one \ac{FRB} per pixel). We find that the measurement of $C_\ell^{Dg}$ is consistent when drawing \acp{FRB} uniformly over the sky and when drawing them based on SFR. 





\subsubsection{Optical followup selection effects}
Redshifts can only be obtained for \acp{FRB} if they are associated with a galaxy that is bright enough to be observable. \acp{FRB} originating from very dim or even ``unseen hosts" \citep[see eg.][]{Marnoch_2023} then will be selected against in any real sample of localized \acp{FRB} with redshift measurements. 

As done in Section \ref{subsec:FRB_Population_Model}, we test the hypothesis that the cross correlation is robust to selection effects that bias against detecting \acp{FRB} in dim hosts directly with our simulation. Although the current magnitude limit for spectroscopic redshifts extends beyond $m_g\sim21$ \citep{cosmos}, we consider a conservative apparent magnitude cutoff and remove all \acp{FRB} in hosts galaxies with $m_g > 20.7$. Our results are shown in \cref{fig:mg}--as expected, the cross-correlation signal is unaffected by the magnitude selection effect on the \ac{FRB} host galaxy.

\subsubsection{Host DM}
\label{sec:hostdm}
Measurements of the \ac{FRB} \ac{DM} include a component from the FRB's host galaxy and local environment, typically of the order of tens to a few hundreds of pc cm$^{-3}$ \citep{Cordes_2022}. For studies that are trying to measure the electron density of the IGM or intervening halos \citep{Connor_ravi_2022,Connor_2024, Sharma_2025}, the \ac{DM} contribution from the host galaxy is a nuisance term that is typically forward modeled. This is further complicated by the fact that the host \ac{DM} distributions of \acp{FRB} are likely to be redshift dependent \cite{Medlock_2025}. One advantage of FRB-DM galaxy cross-correlations is that the host \ac{DM} of the \ac{FRB} is not expected to correlate with the foreground galaxies, and hence the problem of assuming a functional form for the host \ac{DM} contribution is avoided. 

As we discussed above, the resolution of our electron density bins is insufficiently fine to resolve host \acp{DM}. This allows us to verify the assumption that the \ac{FRB} host galaxy \ac{DM} contribution to the cross-correlation is negligible by artificially injecting a host \ac{DM} on top of the \ac{DM} from the ray tracing. To draw the host \ac{DM} values, we use the log-normal parameterization from \cite{Shin_2023}:
\begin{equation} \label{eq:DM_host}
    P(\mathrm{DM}_\text{host}) = \frac{1}{\mathrm{DM}_\text{host}} \frac{1}{\sigma \sqrt{2\pi}} \exp \left[-\frac{(\ln \mathrm{DM}_\text{host} - \mu)^2}{2\sigma^2} \right]
\end{equation}
with best-fit values $\mu = 1.93 / \log_{10} e$ and $\sigma = 0.41 / \log_{10} e$. In \cref{fig:hostDM}, we compare the cross-correlation measurement without the injected host \ac{DM} and with the injection. While the host \ac{DM} contribution increases the variance of the cross-correlation measurement, it does not introduce a bias.  


\subsection{Propagation selection effects}
\label{sec:propagation}
Now we turn to selection effects on observational properties of \acp{FRB} that also arise from propagation effects, namely scattering and DM. Although the host properties of an \ac{FRB} contribute to both the observed scattering and DM, we ignore those effects here since we have demonstrated that the cross correlation is robust to the properties of the \ac{FRB} host in Section \ref{sec:host_selection_effects}.

\subsubsection{DM-dependent selection effects}

To model the DM-dependent selection effects, we use a log-normal probability function based off of the selection function in \cite{2021ApJS..257...59C}
\begin{equation} \label{eq:DM_sfunc}
P_\text{det}(\mathrm{DM}) = \exp \left\{-\frac{2}{3}[\log_{10}(a \cdot \mathrm{DM}) - 3]^2 \right\}
\end{equation}
where \ac{DM} is in units of $\mathrm{pc}\,\mathrm{cm}^{-3}$ and $a$ is a free scaling parameter. The value $a = 1$ roughly corresponds to the DM-dependent selection effect found in \cite{2021ApJS..257...59C}, while $a=2$ very roughly scales it down to the \ac{FRB} redshift range in this study, $z_\text{host} < 0.4$. We compare our results when applying this cut with $a=2$ to the fiducial case in \cref{fig:DM_sfunc1}, and find that the cross-correlation is largely unaffected.

We also test two additional \ac{DM} dependent selection functions that are more aggressive: one using \cref{eq:DM_sfunc} with  $a=5$, and another as a stepwise function cutting all \acp{FRB} with DM$>600$ pc/cm$^{-3}$. The stepwise function is motivated by the limits on realtime memory buffers that can only hold a fixed amount of data and therefore a limited dispersive sweep while \acp{FRB} are searched for in backend systems (e.g. $\sim$ 1000 pc/cm$^{-3}$ for CHIME \citep{Michilli_2021}). 

Our results are shown in \cref{fig:DM_sfunc}. We find that a sufficiently aggressive \ac{DM} dependent selection effect can significantly reduce the amplitude of the cross-correlation signal. Removing the 5\% of \acp{FRB} with the largest \acp{DM} results in the cross-correlation amplitude dropping below $50\%$ at angular scales smaller than $\ell>1000$ (corresponding roughly to physical sizes smaller than $\sim$ 3 Mpc at z=0.35). This implies that the majority of the signal on the scales of galaxy clusters is contained in top 10\% of \ac{FRB} \ac{DM} contributions, since those trace the largest over-densities. 

\begin{figure}[t]
    \includegraphics[width=\columnwidth]{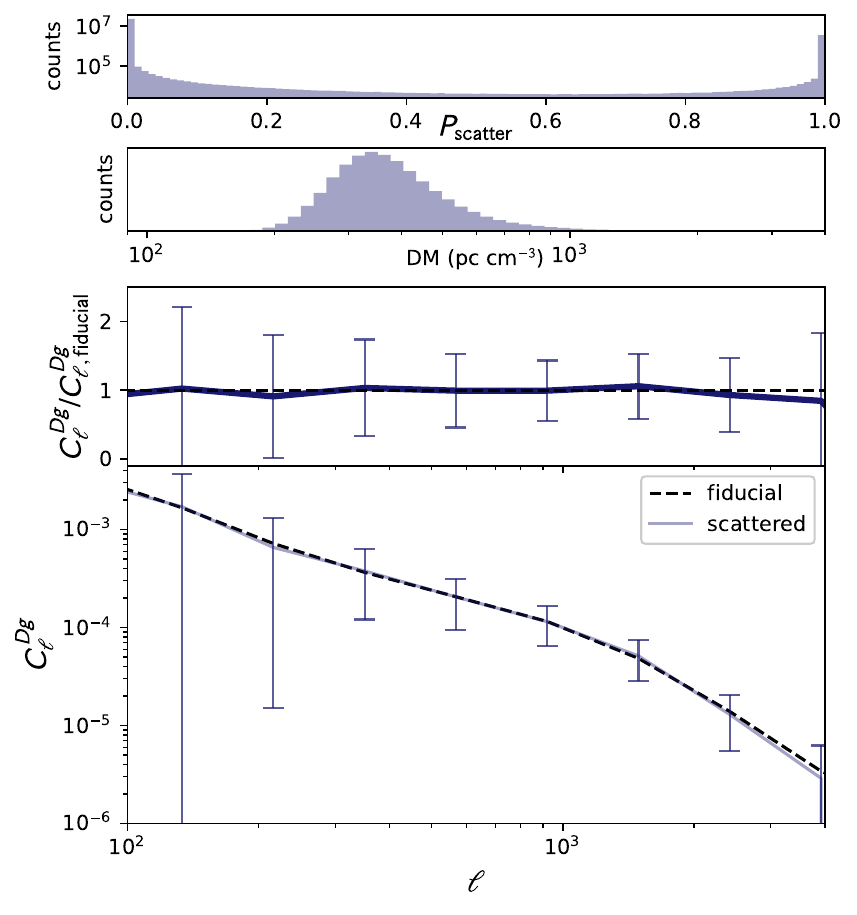}
    \caption{Topmost panel: histogram of probabilities that an \ac{FRB} in our simulation is excluded in the cross-correlation due to the scattering selection function presented in Equation \ref{eq:scatter_sfunc}. Second panel: change in \ac{DM} distribution of \ac{FRB} sample after applying scattering selection effect. Bottom two panels: The lowermost panel shows the cross-correlation amplitudes after applying scattering selection effects (lavender) versus the fiducial sample (dotted magenta), with the ratios shown in the middle panel. Scattering from intervening haloes does not significantly affect of the \ac{DM}-galaxy cross-correlation for our \ac{FRB} sample. }
    \label{fig:scatter}
\end{figure}

\subsubsection{Scattering}
If scattering preferentially removes \acp{FRB} with large electron overdensities along the sightline, it is expected to decrease the amplitude of the cross-correlation signal.  

We model the probability that an \ac{FRB} is not detected due to scattering from intervening galaxies with a simple bell curve with a characteristic time-scale of $1\,\mathrm{ms}$:
\begin{equation} \label{eq:scatter_sfunc}
    P_{\rm{scatter}}(\tau)= 1-2^{-\left( \frac{\tau}{1\,\mathrm{ms}} \right)^2} 
\end{equation}
We simulate scattering timescales for our \acp{FRB} using a simplified version of the cloudlet model based on the work presented by \cite{Ocker_2021}. We stress that while the purely-geometrical scattering model we choose to adopt here is very simplified, the purpose of this analysis is to demonstrate the extent to which scattering can affect the observed measurement of $C_\ell ^ {Dg}$, and we leave more realistic modeling of scattering (e.g. incorporating gas temperature) in ray tracing simulations to future work. 

We calculate the scattering timescale contribution for an \ac{FRB} observed at frequency $\nu$ with an intervening galaxy halo at redshift $z_\ell$ and impact parameter $b_{\perp}$ to be
\begin{equation}
 \label{eq:tau_scatter}
    \tau = 0.3\,\rm{ms}\,\frac{G}{(1+z_\ell)^3(\frac{\nu}{\rm{1\, Ghz}})^4} 2^{-\left( \frac{b_{\perp}}{4\,\mathrm{kpc}}\right)^2}
\end{equation}
which is based on the simulated curves presented in Figure 2 of \cite{Ocker_2021}. G, also known as the ``geometric leverage factor", takes the form
\begin{equation}
    G_g = \frac{2 d_{fg}d_{go}}{d_{fo}L}
\end{equation}
where $d_{fg},d_{go},d_{fo}$ are the FRB-galaxy, galaxy-observer, and FRB-observer distances, respectively, and $L$ denotes the thickness of the scattering medium. We take $L$ to be 1 kpc \citep{Cordes_2022}. We take our observations to be at $\nu=600$Mhz based on the central observing frequency of CHIME, and the total scattering timescale $\tau$ is given by the sum over all intervening galaxy contributions.  

\cref{fig:scatter} shows the cross-correlation with the scattering model applied. We find that scattering due to intervening halos does not significantly affect the cross-correlation signal. However, scattering from intervening material is expected to increase at higher redshifts due to the geometric leverage $G$ \citep{Ocker_2021}. As such, while the cross-correlation appears robust to scattering for our local sample ($z<0.4$), this may not be the case for a catalog of high redshift \acp{FRB}, which requires further investigation in future studies.

\section{Discussion and Conclusion} 
\label{sec:conclusion}
In this work, we have investigated the impact of expected observational selection effects on dispersion--galaxy angular cross-correlations using ray tracing simulations with the Illustris TNG300-1 simulation. Our analysis focused on a local sample of 3000 \acp{FRB} within the redshift range 0.3 $\leq $ z $\leq$ 0.4 cross-correlated with foreground galaxies in a redshift range of 0.2 $\leq $ z $\leq$ 0.3.

Limitations of our simulation framework prevent us from matching our simulated survey parameters to realistic surveys; we can only simulate patches of the sky 6 degrees in extent and ray trace up to a maximum redshift of 0.4. In contrast, CHIME/FRB surveys over half the sky \citep{2018ApJ...863...48C} with its \acp{FRB} extending to redshift 1 \citep{Shin_2023}. As such, our goal is not to precisely quantify the selection effects for any given survey, but to get a first sense of what selection effects are likely to be important. This will provide guidance in planning figure surveys, which will ultimately need to perform simulations matched to their survey  to interpret results. Such larger-scale simulations should be feasible in the future using our ray-tracing framework an larger-box simulations such as MilleniumTNG \citep{MTNG_2023}.

A further limitation is the resolution at which we grid the electron field, limited to 500\,kpc/$h$ pixels with the computational resources available in the JupyterLab workspace provided by IllustrisTNG. This artificially cuts off the electron power spectra at scales \changed{$k \lesssim 6 \,h$\,Mpc$^{-1}$}, and suppresses the power spectrum with the pixel window function at marginally lower $k$. However, it is not expected that extragalactic electrons are strongly clustered on these scales due to feedback processes. In our study, we consider angular scales $\ell < 4000$, which corresponds to physical scales $k < 3.7\,h$ Mpc$^{-1}$ at the galaxy redshift plane. Furthermore, our study focuses not on the magnitude of power spectrum itself, but on the impact of selection effects, and there is no reason to suspect that the latter is strongly resolution dependent.

We demonstrated that these cross-correlations are robust to a large swath of potential systematic errors. These include all effects confined to the host galaxy plane, such as variations in \ac{FRB} host galaxy properties, host \ac{DM} contributions, and optical follow-up selection effects biased against \acp{FRB} with dim galaxy hosts. This result is perhaps expected: the power of cross-correlation studies, in general, are that they are robust to effects confined to only one of the tracers.

We’ve conducted this analysis assuming properties are independent, although in practice this is complicated by the fact that some \ac{FRB} observables (e.g. host \ac{DM} and host luminosity) may be correlated in a complicated way. However, we maintain that this can be expected to have a very small effect since observables that are “local” to the \ac{FRB} are not correlated with the \ac{DM} overdensities and underdensities within the plane of the galaxies that are used in the cross correlation. That is to say the only \ac{FRB} selection effects that can \emph{bias} the power spectrum measurement are \ac{DM} dependent selection effects and other selection effects on “propagation” observables that have a non-zero covariance with the \ac{DM} within the foreground galaxy plane (such as scattering).

A priori, selection biases that related to propagation effects are more concerning, since they correlate with the foreground galaxies. Scattering was initially concerning, since it provides a mechanism by which the lines of sight that pass closest to foreground galaxies are lost. However, in our simulations, which employ a simple model for both scattering and the selection bias against it, we observe no effect on the cross-power spectrum. While scattering due to gas external to galactic discs is highly uncertain, within our current understanding it appears not to be a concern for large-scale structure studies at low $(z<0.4)$ redshifts.

As such, of the systematic errors we have considered, selection bias against \ac{DM} itself is the sole concern for dispersion--galaxy cross-correlations. The magnitude of the bias depends on the details of the selection function. Roughly scaling CHIME's \ac{FRB} selection function to the \acp{DM} in our simulations, we expect the cross-power to be suppressed by order 10\%, worsening on small angular scales. Yet more pernicious is a sharp cutoff in a survey's ability to detect or localize high DMs, such as is present in CHIME's ability to capture baseband data, a requirement for VLBI localization with the Outriggers \citep{outriggers_overview_2025}. We find that excluding the 10\% most-dispersed \acp{FRB} reduces the amplitude of the cross-correlation signal by a factor of two on angular scales of approximately $0.1^\circ$ (corresponding to 
$\sim$ Mpc scales at our simulated galaxy redshift range).

There are a number of ways that future measurements of the dispersion--galaxy cross-power spectrum could overcome the bias from DM-dependent selection effects. The selection effects could be simulated---using the framework developed here but with survey parameters matched to the measurement---and accounted for in any interpretations. Alternately, it may be possible to develop an analytic model.

Another strategy is to reduce or eliminate the selection effects in the first place. The sharp cutoff bias in CHIME could be overcome via memory upgrades for the baseband systems, including the Outriggers. The \ac{FRB} detection selection function, coming mainly from dispersive smearing reducing S/N, should be less severe for instruments operating at higher frequencies or with better frequency resolution. Even without instrument changes, a bias-free subsample could be created by selecting detected events above a fixed fluence, which is invariant to dispersion.

Our findings suggest that cross-correlation techniques remain a promising method to probe the distribution of baryons using \ac{FRB} DMs, but selection effects to dispersion must be properly accounted for to avoid biased measurements. As next-generation radio telescopes such as CHORD, \changed{DSA-2000}, and BURSTT begin to collect larger and more precise samples of FRBs, our results provide a framework for understanding and mitigating selection biases that may impact \ac{FRB} surveys aimed at probing the large-scale distribution of baryons.


\section{Data Availability}
All code, including a \ac{FRB} ray-tracing package \texttt{illustris\_frb} and scripts for processing Illustris data and running experiments, is available on Github\footnote{\url{https://github.com/aqcheng/illustris_frb}}, along with the notebook for generating all data and figures in this paper.

\section*{acknowledgments}
\begin{acknowledgments}
We thank Kaitlyn Shin for their helpful comments on this paper. We would also like to thank Ryan Raikman, Ralf Konietzka, Liam Connor, and Calvin Leung for conversations that have helped inform this paper. \\
A.Q.C. was partially supported by the MIT UROP program for the majority of this work, and is currently supported by the Lowell Wood Endowed Fellowship of the Fannie and John Hertz Foundation. K.W.M. holds the Adam J. Burgasser Chair in Astrophysics and received support from NSF grant 2018490.
\end{acknowledgments}

\appendix

\section{Power spectrum estimator derivation} \label{app:estimator}

In this section, we show that the \ac{OQE} (\cref{eq:oqe,eq:fisher}) reduces to the naive estimator (\cref{eq:naive_estimator_flat_binned}) in the case of complete sky coverage, i.e. when $\mathbf{D}^{-1}$ and $\mathbf{G}^{-1}$ are proportional to the identity.
First, let us compute the derivative of the covariance matrix \dcov. Recall from \cref{eq:DFT} that the 2D \ac{DFT} over a flat square sky patch of $N \times N$ pixels for some field $f_{\vec{\theta}}$ is
\begin{align}
\tilde{f}(\vec{\ell}) &= \frac{A}{N}\sum_{\vec{\theta}} Q_{\vec{\ell} \vec{\theta}}f_{\vec{\theta}} \\
f(\vec{\theta}) &= \frac{N}{A} \sum_{\vec{\ell}} Q^\dagger_{\vec{\theta} \vec{\ell}}\tilde{f}_{\vec{\ell}}
\end{align}
where we have defined the orthonormal operators
\begin{align} 
    Q_{\vec{\ell} \vec{\theta}} &= \frac{1}{N} e^{-i \vec{\ell} \cdot \vec{\theta}} \\
    Q^\dagger_{\vec{\theta} \vec{\ell}} &= \frac{1}{N} e^{i \vec{\ell} \cdot \vec{\theta}}
\end{align}
where $\vec{\theta}$ is the position vector in the 2-dimensional configuration space and $\vec{\ell}$ is the wavevector.

Therefore, \dcov can be computed as
\begin{equation} \label{eq:derive_dcov}
\begin{split}
    (C_{, \lambda})_{\vec{\theta} \vec{\theta}^\prime} = \frac{\partial C_{\vec{\theta} \vec{\theta}^\prime}}{\partial C^{Dg}_{\lambda}} & = \frac{\partial \left \langle d_{\vec{\theta}} g^*_{\vec{\theta}^\prime} \right \rangle}{\partial C^{Dg}_{\lambda}}  \\
    & = \frac{N^2}{A^2}\frac{\partial}{\partial C^{Dg}_{\lambda}} \sum_{\vec{\ell} , \, \vec{\ell}^\prime}  Q^\dagger_{\vec{\theta} \vec{\ell}} \left \langle \tilde{d}_{\vec{\ell}} \hbox{ } \tilde{g}^*_{\vec{\ell}^\prime} \right \rangle  Q_{\vec{\ell}^\prime \vec{\theta}^\prime}  \\
    & = \frac{N^2}{A} \frac{\partial}{\partial C^{Dg}_{\lambda}} \sum_{\vec{\ell}}  Q^\dagger_{\vec{\theta} \vec{\ell}} \hbox{ } C^{Dg}_\ell  Q_{\vec{\ell} \vec{\theta}^\prime} \\  
    & = \frac{N^2}{A} \sum_{\vec{\ell} \in \lambda} Q^\dagger_{\vec{\theta} \vec{\ell}} \hbox{ } Q_{\vec{\ell} \vec{\theta}^\prime},
\end{split}
\end{equation}
where we substituted the definition of the cross-correlation \cref{eq:ang_spec_def_flat} on the third line, and substituted $C_\ell^{Dg} \approx C_\lambda^{Dg}$ for $|\vec{\ell}| \in \lambda$ in the last line. 
With complete sky coverage, the \cref{eq:oqe} reduces to
\begin{equation} \label{eq:simplified_oqe}
    C_\lambda^{Dg} = \sum_{\lambda^\prime} \trace[\dcov \dcovp]^{-1} \mathbf{g}^\dagger \dcovp \mathbf{d}
\end{equation}
With the result from \cref{eq:derive_dcov}, we can compute the first term as 
\begin{equation} \label{eq:oqe_trace_term}
\begin{split}
    \trace[\dcov \dcovp] &= \frac{N^4}{A^2}  \sum_{|\vec{\ell}| \in \lambda} \sum_{|\vec{\ell}^\prime| \in \lambda^\prime} \sum_{\vec{\theta}, \vec{\theta}^\prime}  Q_{\vec{\ell} \vec{\theta}} \,Q^\dagger_{\vec{\theta}\vec{\ell}^\prime} Q_{\vec{\ell}^\prime \vec{\theta}^\prime} Q^\dagger_{\vec{\theta}^\prime\vec{\ell}} \\
    &= \frac{N^4}{A^2} \sum_{|\vec{\ell}| \in \lambda} \sum_{|\vec{\ell}^\prime| \in \lambda^\prime} \delta_{\vec{\ell} \vec{\ell}^\prime} = \frac{N^4}{A^2} N_\lambda,
\end{split}
\end{equation}
where $N_\lambda$ is the number of modes in the bandpower $\lambda$. The remainder of \cref{eq:simplified_oqe} is computed as
\begin{equation} \label{eq:oqe_numerator_term}
\begin{split}
    \mathbf{d}\tp \dcovp \mathbf{g} &= \frac{N^2}{A}\sum_{|\vec{\ell}| \in \lambda} \sum_{\vec{\theta}, \,\vec{\theta}^\prime}  d_{\vec{\theta}} \, Q^\dagger_{\vec{\theta} \vec{\ell}} Q_{\vec{\ell} \vec{\theta}^\prime}\,g_{\vec{\theta}^\prime} \\
    &= \frac{N^4}{A^3}\sum_{|\vec{\ell}| \in \lambda} \tilde{d}^*_{\vec{\ell}} \,\tilde{g}_{\vec{\ell}}
\end{split}
\end{equation}
Thus \cref{eq:simplified_oqe} is equivalent to
\begin{equation}
    \hat{C}_\lambda^{Dg} = \frac{1}{A\,N_\lambda} \sum_{|\vec{\ell}| \in \lambda} \tilde{d}^*_{\vec{\ell}} \,\tilde{g}_{\vec{\ell}},
\end{equation}
exactly the naive estimator in \cref{eq:naive_estimator_flat_binned}.

Finally, we show that the \ac{OQE} in our implementation is on average equivalent to the so-called pseudo-$C_\ell$ estimator \citep{Alonso_2019}, which is the naive estimator scaled by the available sky fraction:
\begin{equation} \label{eq:simple_qe}
   \langle \hat{C}_\lambda^{Dg} \rangle = \frac{1}{f_{sky}}\frac{1}{A N_\lambda} \sum_{\vec{\ell} \in \lambda} \left\langle \left(\tilde{d}^m_{\vec{\ell}}\right)^* \tilde{g}^m_{\vec{\ell}} \right\rangle,
\end{equation}
With a general window function, the term $\mathbf{d}\tp \dcovp \mathbf{g}$ stays the same since $\icovd \mathbf{d}^m = \mathbf{d}^m$ and $\icovg \mathbf{g}^m = \mathbf{g}^m$, i.e. applying a mask to a masked field does not change it. The trace term \cref{eq:oqe_trace_term}, however, is generalized to
\begin{equation}
\begin{split}
    \trace[\dcov \icovg \dcovp  \icovd] &= \frac{N^4}{A^2}\sum_{|\vec{\ell}| \in \lambda} \sum_{|\vec{\ell}^\prime| \in \lambda^\prime} \sum_{\vec{\theta}, \vec{\theta}^\prime} Q^\dagger_{\vec{\theta} \vec{\ell}} ~ Q_{\vec{\ell} \vec{\theta}^{\prime}} \,(G^{-1})_{\vec{\theta}^{\prime}} Q^\dagger_{\vec{\theta}^{\prime} \vec{\ell}^\prime} ~ Q_{\vec{\ell}^\prime \vec{\theta}} (D^{-1})_{\vec{\theta}} \\
    &= \frac{N^4}{A^2} \sum_{|\vec{\ell}| \in \lambda} \sum_{|\vec{\ell}^\prime| \in \lambda^\prime} (\mathbf{Q} \icovd \mathbf{Q}^\dagger)_{\vec{\ell}\vec{\ell}^\prime}(\mathbf{Q} \icovg \mathbf{Q}^\dagger)_{\vec{\ell}^\prime\vec{\ell}}.
\end{split}
\end{equation}
In our case, we have complete sky coverage for galaxies ($\icovg = \mathbf{I}$), while \acp{FRB} occupy a random fraction $f_{sky}$ of pixels in the sky region. Therefore, $\langle \icovg \rangle = f_{sky}\mathbf{I}$ and 
\begin{equation}
    \langle \trace[\dcov \icovg \dcovp  \icovd] \rangle = \frac{N^4}{A^2} N_\lambda f_{sky}.
\end{equation}
Taking the expectation value of $C_\lambda^{Dg}$ in \cref{eq:oqe} and assuming that the survey window is uncorrelated with the \ac{DM} field then yields the result in \cref{eq:simple_qe}.

\section{Simulation sky patch selection} \label{app:skypatch}

We consider two conditions in our selection of good sky patches: (1) no ray within the region will intersect itself, and (2) no two rays within the region will intersect the same patch of the simulation box. 

The first problem of the self-intersecting ray is manifest in any field with periodic boundary conditions. For example, naively following a ray parallel to any axis would result in repeating the same exact path through the simulation box, once we enforce periodic boundary conditions. A ray to a \ac{FRB} at $z=0.4$ for a simulation box of size 300 cMpc would cross the same structures $5$ times.

To identify which sky directions are problematic up to $z=0.4$, recall that periodic boundary conditions identifies any point $\vec{x}$ within the simulation box with points $\vec{x} + b(k, m, n)$, where $b$ is the simulation box size and $k, m, n$ are integers. Now, consider the vector corresponding to the ray, $\vec{v} = \vec{x}_{FRB} - \vec{x}_0$.  Condition (1) is equivalent to the condition that $\vec{x}_0 + b(k, m, n)$ does not lie anywhere along the ray for all $k, m, n$. Therefore, we can find all problematic sky directions by computing all $|b(k, m, n)| < |\vec{v}| = \chi(z=0.4)$, where $\chi$ is the comoving distance. Because our simulation box is gridded into bins of size $500\,\mathrm{ckpc}/h$, each of these sky directions excludes a circular sky area of radius $\theta(z) = \frac{500/(1+z)\,\mathrm{kpc}/h}{D_A(z)}$ evaluated at $z=0.4$, where $D_A$ is the angular diameter distance. In a flat cosmology, this is a radius of $\theta(z) = \frac{500\,\mathrm{ckpc}/h}{\chi(z)}\approx 5 \cdot 10^{-4}\,\mathrm{rad}$ at $z=0.4$.

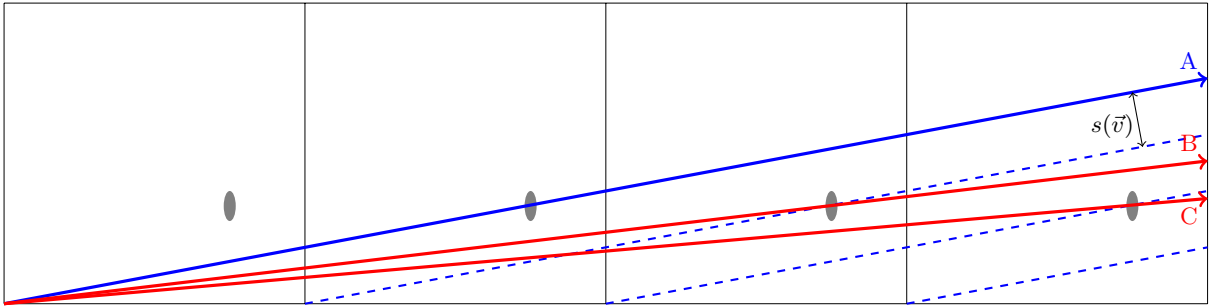
\begin{figure}[h]
\centering
\begin{tikzpicture}
\draw[step=4cm,very thin] (0,0) grid (16,4);
\foreach \x in {3,7,11,15}
    \fill[gray] (\x,1.3) ellipse (0.08cm and 0.2cm);
\draw[blue,very thick,->] (0,0) -- (16, 3) node[anchor=south east] {A}; 
\draw[blue, thick,dashed] (4,0) -- (16, 9/4);
\draw[blue, thick,dashed] (8,0) -- (16, 1.5);
\draw[blue, thick,dashed] (12,0) -- (16, 0.75);
\draw[red,very thick,->] (0,0) -- (16, 1.9) node[anchor=south east] {B}; 
\draw[red,very thick,->] (0,0) -- (16, 1.4) node[anchor=north east] {C}; 
\draw[black,<->] (15,45/16) -- (15.13585, 2.08797) node[anchor=south east] {$s(\vec{v})$};
\end{tikzpicture}
\caption{Parallel sightlines can intersect each other in different periodic boxes. Here, nearby sightlines B and C (red) intersect the images (dashed blue lines) of sightline A in adjacent boxes. Notably, a large intervening structure (e.g. a galaxy cluster) can yield spurious spatial patterns, as sightlines A, B, and C would have \ac{DM} contributions from the same structure (oval), whereas sightlines in between them would not.}
\label{fig:condition2}
\end{figure}

The second problem is the intersection of \textit{nearby} sightlines, e.g. a grid of \acp{FRB} within a small sky patch. This can occur when stacking periodic boxes; see \cref{fig:condition2}. In order to avoid this, the sightline $\vec{v}$ must be sufficiently far apart from its images $\vec{v} + b(k,m,n)$, in particular its adjacent images (i.e. $k, m, n \in \{0, 1\})$. An ``image" here refers to any identical path to the sightline as enforced by the periodic boundary condition. \cref{fig:condition2} visualizes this issue: sightline A (solid blue) must be sufficiently far apart from its images (dashed blue) such that the other sightlines in the same sky region do not intersect them.

Let $\vec{u}$ be any $b(k, m, n)$ with $k, m, n \in \mathbb{Z}$. The smallest distance between two rays $\vec{v}$ and $\vec{v} + \vec{u}$ is given by $\hat{v} \times \vec{u}$, where $\hat{v} = \frac{\vec{v}}{|\vec{v}|}$ is the unit vector corresponding to $\vec{v}$ and uniquely identifies a sky direction. Thus for any given sky direction $\hat{v}$, we can compute the distance to the closest adjacent image
\begin{equation}
    s(\hat{v}) = \min_{\vec{u}} \left| \hat{v} \times \vec{u} \right|
\end{equation}
Therefore, for a given redshift $z$ at which the \acp{FRB} will be placed, the selection of a ``good" sky patch of angular size $\theta$ is the requirement
\begin{equation}
    s(\hat{v}) > \theta \,\chi(z).
\end{equation}
Applying both of these conditions to yields the map of good sightlines, as shown by the white regions in \cref{fig:regions} for $\theta=0.1\,\mathrm{rad}$ and $z=0.4$. The requirement (2) that nearby rays cannot intersect in adjacent boxes is extremely restrictive. Furthermore, the ``good" sky areas shrink when we ray trace to greater distances, as the ray crosses more periodic boxes. This restriction sets a limit for the sky region sizes and \ac{FRB} redshifts we can explore within IllustrisTNG. It is for this reason that we choose a sky patch size of $0.1$ rad and a maximum redshift of $z=0.4$.

\bibliography{bib}{}
\bibliographystyle{aasjournal}

\end{document}

%% file: acronyms.tex
\acrodef{DM}{dispersion measure}
\acrodef{FRB}{Fast Radio Burst}
\acrodef{SFR}{star formation rate}
\acrodef{OQE}{optimal quadratic estimator}
\acrodef{IGM}{intergalactic medium}
\acrodef{SHT}{spherical harmonic transform}
\acrodef{DFT}{discrete Fourier transform}

%% file: commands.tex
\newcommand{\changed}[1]{#1}

\newcommand\DM{\ensuremath{\mathrm{DM}}\xspace}
\newcommand\rad{\ensuremath{\mathrm{rad}}\xspace}
\newcommand\ClDg{\ensuremath{C_\ell^{Dg}}\xspace}
\newcommand\ckpch{\ensuremath{\mathrm{ckpc}/h}\xspace}

\newcommand\cov{\ensuremath{\mathbf{C}}\xspace}

\newcommand\icovd{\ensuremath{\mathbf{D}^{-1}}\xspace}
\newcommand\icovg{\ensuremath{\mathbf{G}^{-1}}\xspace}
\newcommand\dcov{\ensuremath{\mathbf{C}_{, \lambda}}\xspace}
\newcommand\dcovp{\ensuremath{\mathbf{C}_{, \lambda^\prime}}\xspace}
\newcommand\tp{\ensuremath{^\mathrm{T}}\xspace}
\newcommand\fisher{\ensuremath{F_{\lambda\lambda^\prime}\xspace}}

\newcommand\trace{\ensuremath{\mathrm{Tr}}\xspace}